# Imaging the indirect dissociation dynamics of temporary negative ion: $N_2O^-$ → $N_2 + O^-$

Lei Xia, Bin Wu, Hong-Kai Li, Xian-Jin Zeng, and Shan Xi Tian*

*Hefei National Laboratory for Physical Sciences at the Microscale and Department of Chemical Physics, University of Science and Technology of China, 96 Jinzhai Road, Hefei, Anhui 230026, China*

*Corresponding Author. sxtian@ustc.edu.cn

ABSTRACT

We reported an imaging study of the dissociation dynamics of temporary negative ion $N_2O^-$ formed in the low-energy electron attachment, $e^- + N_2O \rightarrow N_2O^- \rightarrow N_2 + O^-$. With the help of *ab initio* molecular dynamics calculations, the evolution of momentum distributions of the $O^-$ fragment in terms of the electron attachment energy is identified as the result of a competition between two distinctly different indirect pathways, namely, climbing over and bypassing the energy ridge after the molecular structure bending. These two pathways prefer leaving the $N_2$ fragment at the high vibrational and rotational states, respectively.

Unimolecular dissociation provides a suitable vessel for investigation of chemical reaction dynamics.[1] Direct and indirect processes represent the two major classes of the molecular dissociation induced by photoabsorption or collision. In the direct case, the fragmentation proceeds too fast to develop internal energy of the target molecule; in the indirect case, the dissociation usually experiences a metastable state which has a sufficiently long time to allow the internal energy redistribution. For the latter, the final state distributions of the fragments no longer reflect the initial state wavefunction like they do in the direct dissociation. The details of the indirect dissociation are, even in a black box, rather difficult to be revealed, unless the final states of all fragments can be measured precisely and the various dissociation pathways on the potential energy surface (PES) at the intermediate state can be theoretically simulated.[2] Electron-molecule resonance, also know as temporary negative ion (TNI), is a typical intermediate resonant state formed in the low-energy electron attachment to molecule and may decay via dissociation to a negatively charged ion and one or more neutral fragments.[3] Due to the very short lifetime of TNI (from microseconds to femtoseconds or less, depending on the resonance types[3]), most dissociations, even for the large flexible molecules, are believed to be the direct processes.[4-6] Here we present an imaging study of the dissociation dynamics of a triatomic TNI $N_2O^-$, $N_2O^- \rightarrow N_2 + O^-$, by the angular-resolved detection of the $O^-$ fragment. Two modes of the $O^-$ momentum distributions confidently provide an evidence of the indirect dissociation: after the structural bending of the linear molecule, two distinctly different dissociation pathways, namely, climbing over and bypassing the energy ridge on the PES of $N_2O^-$, compete with each other.

The full-collision dynamic picture of the dissociative electron attachment (DEA) via the predissociation-state TNI, especially for a polyatomic molecular target, is still a challenge both in the experiments and theoretical calculations.[5] In the low-energy (less than several eV) electron attachment to $N_2O$, several series of vibrational excitations of N-O bond stretching ($\upsilon_1$), N-N-O bond bending ($\upsilon_2$), and N-N bond stretching ($\upsilon_3$) modes were observed around the $N_2O^-$ $^2\Pi$ shape resonant state;[7,8] a diffuse band including two peaks (0.7 eV and 2.4 eV) in the $O^-$ yield spectra of the DEA process[7,9] can also be attributed to this resonant state on the basis of the recent theoretical calculations.[10] However, a long-term argument about the coexistence of a $^2\Sigma$ resonant state in this energy range arose from the measurements of the angular-resolved differential cross sections of the $O^-$ ion in a limited angular range (30° to 130°).[7,11] It is also puzzling that the kinetic energies of the $O^-$ ion were quite large even at the low electron incident energies.[8] Herein, these misconceptions will be clarified.

The $O^-$ momentum distributions for the DEA process of $N_2O$ have been measured at the electron energies of 0.70, 1.90, 2.25, and 2.60 eV by using our newly developed anion velocity image mapping apparatus.[12] As shown in Fig. 1, the time-sliced images of the $O^-$ momentum clearly show the pattern evolution, namely, from the half-rings (Fig. 1a) to the tetrad-petal-like images (Figs. 1c and 1d). Their common features are: the stronger backward scattering, the nodes along the *z* axis, and the nearly same size of the momentum rings. The ring size is determined by the $O^-$ velocity projected on the *x-z* detection plane. Before going to reveal the dissociation dynamics underneath these images, the PES of $N_2O^-$ is plotted within the Jacobi coordinates R and γ (see their definitions in the upright panel of Fig. 2) on the basis of the configuration interaction calculations at CASSCF(11,10) / 6-311G(d) level (more details can be found as the supplementary material). In the formation of $N_2O^-$ at $^2\Pi$ resonant state, the excess electron is captured and occupies at the lowest unoccupied molecular orbital $\pi^*$. Due to the Renner-Teller effect, the $^2\Pi$ resonant state undergoes splitting into $^2A''$ and $^2A'$ states in asymptote if the N-N-O bond bending ($\upsilon_2$) vibration is involved in the electron attachment. No other states could be involved in the DEA processes investigated here.[8-10] In Fig. 2, the black thick line indicates the PES intersection curve at the $^2\Pi$ resonant state, where the N-O bond stretching ($\upsilon_1$) and N-N bond stretching ($\upsilon_3$) vibrations are permitted (the linear molecular structure is still kept). The direct dissociation $N_2O^- \rightarrow N_2 + O^-$ at the $^2\Pi$ resonant state by an elongation of N-O bond may be hindered due to an energy barrier about 0.68 eV (close to the previously speculated value 0.5 eV[9]), but an indirect dissociation is feasible. After the vertical attachment, the linear structure (R = 1.748 Å and $\gamma = 0°$) will be slightly bent ($\gamma \approx 2°$) along the red arrow which points the low energy region then is followed with two distinctly different dissociation pathways. One pathway is named as the 'over-of-the-ridge' path and labeled with the blue arrow, in which the energy barrier along the R axis needs to be overcome. This energy ridge dramatically collapses with the increase of $\gamma$, from 0.68 eV (R = 2.198 Å, $\gamma = 0°$) to 0.08 eV (R = 2.048 Å, $\gamma = 20°$); therefore, the $N_2O^-$ can also go down firstly along the declining slope with the increasing $\gamma$ values, then happens to go through the much lower ridge at a larger $\gamma$ value. This is noted as the 'ridge-bypassing' path and labeled with the yellow arrow. Although these two paths are on the $^2A'$ PES, the over-of-the-ridge path is approximately of the $^2\Pi$ symmetry because it is nearly parallel to the intersection curve of the $^2\Pi$ PES (thick line in Fig.2). These two different pathways are crucial to reveal the evolution dynamics observed in the O¯ momentum images.

The previous experiments showed the O¯ kinetic energies were 0.23 and 0.44 eV at the incident energies 0.612 and 2.5 eV, respectively.[8] The thermodynamic energy threshold 0.21 eV of $N_2O^- \rightarrow N_2 + O^-$ was determined as well.[8] If the residual energy after the dissociation is completely transformed to be the translational energies of the fragments, the maximum kinetic energy of the O¯, $E_{max} = (E_{incident} - 0.21)/(1 + 16/28)$, should be 0.31, 1.08, 1.30, and 1.52 eV for the attachment energies 0.70, 1.90, 2.25, and 2.60 eV, respectively. By integration of the image data, the kinetic energy distributions of the O¯ fragment are plotted in Fig. 3(a). At the different attachment energies, all kinetic energy profiles show the peaks at about 0.33 eV, in accord with the previous results.[8] However, the kinetic energies around 0.33 eV are much lower than the above predictions, implying that the most residual energies should be transformed to be the internal energies of $N_2$ fragment.

The angular-resolved differential cross sections of the O¯ fragment in the kinetic energy range of 0.15 - 0.50 eV are depicted as the solid circles in Figs. 3(b-e). For the DEA of diatomic molecules within the axial-recoil approximation, the momentum distribution of the anion fragment is determined by,[13]

$$\sigma_{anion}(k,\theta,\varphi) \propto \sum_{|\mu|} \left| \sum_{l=|\mu|}^{\alpha} a_{l\mu}(k) Y_{l\mu}(\theta,\zeta) \right|^2 \approx \sum_{|\mu|} \left| \sum_{j=1, l=|\mu|} c_j e^{i\delta_j} Y_{l\mu}(\theta,\zeta) \right|^2 \quad (1)$$

where $a_{l\mu}(k)$ is the attachment-energy ($k$) dependent expansion coefficient and further expressed with partial waves of the different angular momentum $l$ ($l \geq |\mu|$) of the impinging electron, while $Y_{l\mu}$ is the spherical harmonics describing the neutral and TNI states. $|\mu|$ equals to $|\Lambda_f - \Lambda_i|$, representing the difference in the projection of the angular momentum along the internuclear axis for the neutral and TNI. In the formation of a resonant state of TNI, the different influences on each partial wave of the impinging electron by the interaction potential of the target result in the phase lags ($\delta_j$) among these partial waves. In practice, the summation of the finite partial wave terms together with the weight parameter $c_j$ can reproduce the momentum distribution of the anionic fragment for single resonant state.[4,511] The present DEA should be beyond the axial-recoil approximation due to the structural bending ($\gamma \approx 2°$) after the electron attachment. Assuming that the departure between the O⁻ and N$_2$ fragments is approximately along a new recoil axis, the axial-recoil approximation is supposed to be valid for the dissociation of this slightly bending N$_2$O⁻. In this binary dissociation, $\Lambda_f = 1$ ($^2\Pi$, the blue arrow) along the over-the-ridge pathway; while $\Lambda_f = 0$ ($^2$A', the yellow arrow) along the ridge-bypassing pathway. Therefore, the differential cross section of the O⁻ fragment can be rewritten as,

$$\sigma_{anion}(\theta) \propto \left|a_0 Y_{00} + e^{i\delta_1} a_1 Y_{10} + e^{i\delta_2} a_2 Y_{20}\right|^2 + \left|b_0 Y_{11} + e^{i\delta_3} b_1 Y_{21} + e^{i\delta_4} b_2 Y_{31}\right|^2 \quad (2)$$

where the over-the-ridge and ridge-bypassing pathways are represented with the first and second terms in the right. As shown in Figs. 3(b-e), the excellent experimental data fittings using eq. (2) are achieved with the high correlations (> 0.99). The fitted parameters (given as the supplementary material) indicate that the predominant scattering amplitudes are $s\Sigma(^2A') + d\Pi$ at the attachment energies of 0.70 and 1.90 eV, while $s\Sigma(^2A') + p\Pi + d\Pi$ at 2.25 and 2.60 eV. The previous argument about using either $\Sigma$ or $\Pi$ symmetry in fitting the momentum distribution[7,11] now can be clarified satisfyingly.

To elucidate the different roles of each pathway, the momentum distributions at 0.70 and 2.60 eV are reproduced by using the fitted parameters and shown as the broken curves in Figs. 3(b) and 3(e). The backward-scattering features are primarily attributed to the ridge-bypassing pathway (the red broken curves), while the over-the-ridge pathway leads to the significant tetrad-petal of the $O^-$ momentum distribution (the black broken curves). In Fig. 3(f), the branching ratio of these two pathways [$\sigma_{anion}(^2A')/\sigma_{anion}(^2\Pi)$] has been plotted by the data integration over θ (0 ~ 2π) using the fitting parameters. The contribution of the ridge-bypassing pathway [$\sigma_{anion}(^2A')$] overwhelms that of the over-the-ridge pathway [$\sigma_{anion}(^2\Pi)$] at these four attachment energies, although the former decreases quickly from 0.70 eV to 1.90 eV. It is amazing that even at the higher energy 2.60 eV the dissociation $N_2O^- \rightarrow N_2 + O^-$ still prefers the ridge-bypassing path.

According to the momentum and energy conservation principles and supposing that the most $O^-$ anions have the kinetic energy of 0.33 eV (the central value of the kinetic energy distributions in Fig. 3a), the $N_2$ species should carry out the translational energy about 0.19 eV, but have the different internal energies, 0.00, 1.17, 1.52, and 1.87 eV at the attachment energies, 0.70, 1.90, 2.25, and 2.60 eV, respectively. A quite striking question is how these internal energies are participated into, e.g., the high rotational quanta $j$, vibrational quanta $\upsilon$, or both of $N_2$? This is definitely dependent on the dissociation paths or the dynamic trajectories on the PES of Fig. 2.

Due to the extremely low cross sections of DEA, the electron monochromator is still unavailable in the present studies.[12] The energy spread of the incident electron is ± 0.3 eV, thereby the possible rovibrational states of $N_2$ cannot be resolved in the $O^-$ kinetic-energy distribution profiles (see Fig. 3a). We performed the *ab initio* molecular dynamics simulations for the dissociation trajectories of the $N_2O^-$ by using the atom-centered density matrix propagation method [ADMP-B3LYP/6-311G(d), more details can be found as the supplementary material]. As mentioned above, the structural propagation is initiated in the Franck-Condon region for the vertical attachment to the linear $N_2O$. Quasi-classical trajectories starting with two typical initial internal energies 0.01 and 1.00 eV of the TNI can mimic two different DEA processes at the lowest and highest attachment energies. The snapshots of the sample trajectories are selected as eight frames of Fig. 4. At 20 fs, the $N_2O^-$ structure is slightly bent [angle (O-N-N) is about 177°] and the N-O bond breaks. For the low internal energy, see Fig. 4(a), $N_2$ fragment further receives the torque from the N-N-O bond bending. A vector correlation $v \perp j$ is observed, where $v$ is the $O^-$ velocity and $j$ is the rotational moment of $N_2$ (potentially at the higher rotational quanta). For the high internal energy, see Fig. 4(b), the vibrational excited $N_2$ fragment is observed. At 86 fs, $N_2$ and $O^-$ fragments are almost aligned. It is noted that there are still some trajectories exhibiting the highly rotating $N_2$, which is similar to the scenario of the low internal energy simulations. This is in line with our conjecture that the $N_2$ species would like to continuously receive the torque by the molecular bending due to the large PES gradient along the γ axis. Therefore, the ridge-bypassing pathway is still predominant even at the higher energies (see Fig. 3f). The quasi-classical trajectories provide very good qualitative descriptions to the significant differences of these two dissociation pathways. For example, at the incident energy of 1.90 eV, the rotational quanta of $N_2$ could reach $j \approx 68$ [$E_{int}$ =

$j(j+1)h^2/8\pi^2 I$, where the total internal energy $E_{int}$ = 1.17 eV, $I$ is the moment of inertia of $N_2$] at the vibrational ground state $v = 0$, if the dissociation occurs along the ridge-bypassing pathway; or the $N_2$ fragment may be at the vibrational excited state $v \approx 4$ without the rotational excitation, if the dissociation proceeds in the over-the-ridge pathway. In a recent theoretical study, the authors also suggested that the strong rotation-exited $N_2$ fragment could be produced together with $O^-$ ion on the $^2A'$ PES while the dissociation was inhibited on the $^2A''$ PES.[14] Herein, the vibration-excited $N_2$ fragment as the minor is proposed to be produced via the over-the-ridge path on the $^2A'$ PES, i.e., by the N-O bond elongation.

In summary, we reported the convincing experimental evidence about the complicated indirect dissociation dynamics of $N_2O^-$. The product state distribution is heavily related to the state evolution of $N_2O^-$, rather than the neutral state. The present electronic structure calculations and dynamics simulations can provide us for a good qualitative picture of the indirect dissociations, but the competition with electron autodetachment due to the coupling with the electron continuum[15] is ignored. The more sophisticated quantum scattering calculations should be required for the dissociation dynamics of TNI.[15,16] The previous work using anion mass spectrometer proposed that some unanticipated products could be produced via the indirect process of DEA.[16] The present anion velocity imaging spectrometry is a much more promising reaction microscope providing insights into the complicated dynamics of the DEA processes.

This work is supported by MOST (Grant No. 2013CB834602), NSFC (Grant No. 21273213), and FRFCU.

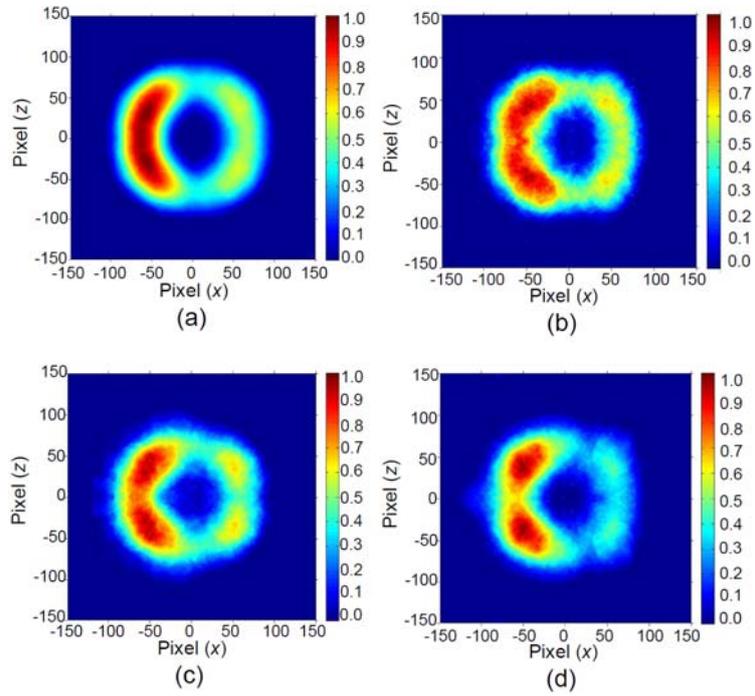

**FIG. 1** Sliced images of the O¯ momentum at the electron attachment energies of 0.70 eV (a), 1.90 eV (b), 2.25 eV (c), and 2.60 eV (d). The electron incident direction is from the left to the right. The anion intensities of the images are normalized respectively.

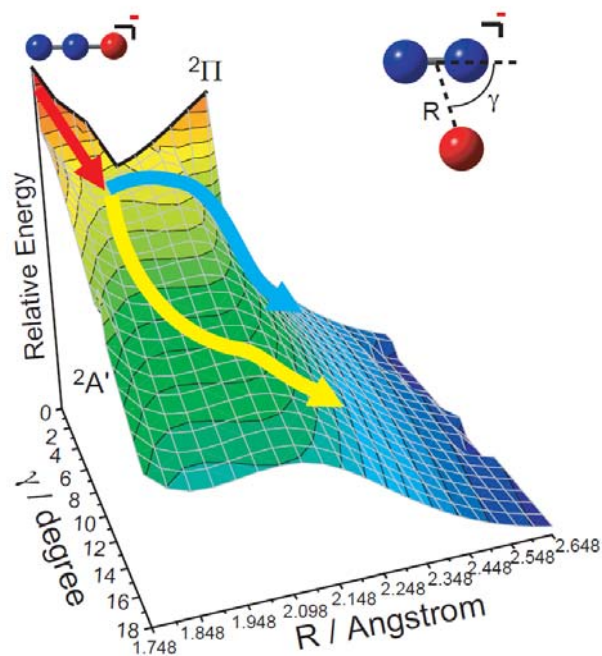

**FIG. 2** Potential energy surface in the Jacobi coordinate. The Jacobi coordinate is defined as the upright panel: R is the distance between oxygen and the midpoint of molecular nitrogen bond and γ is the orientation angle of the $N_2$ bond axis. The red arrow denotes the simultaneous TNI structure bending; The blue and yellow arrows represent the over-the-ridge and the ridge-bypassing, respectively.

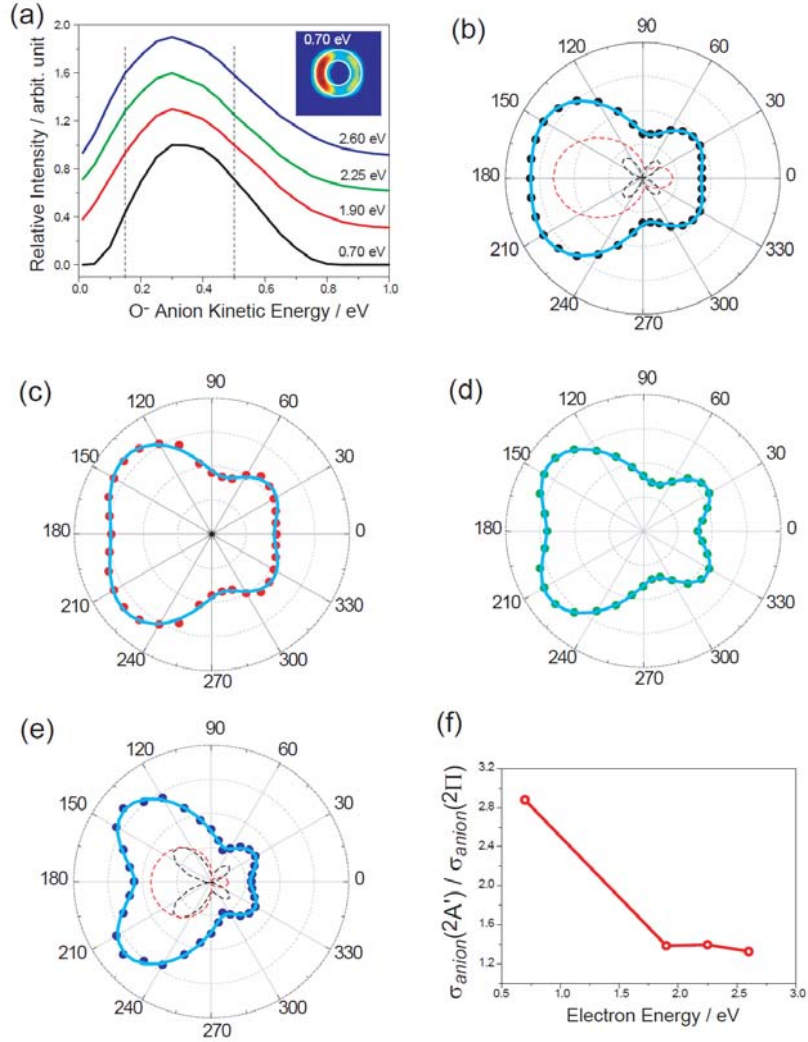

**FIG. 3** (a) The kinetic energy distributions of the O¯ fragment are plotted by integration of ion signals in the selected annular area of the sliced image in Fig. 1. Two vertical broken lines show the higher intensities in the kinetic energy range of 0.15 – 0.50 eV which corresponds to the annular area of the sliced image, for example, at the electron incident energy of 0.70 eV, see the inserted panel. The momentum distributions of the O¯ fragment in the kinetic energy range of 0.15 – 0.50 eV at the different electron energies: (b) 0.70 eV, (c) 1.90 eV, (d) 2.25 eV, and (e) 2.60 eV. The broken curves in (b) and (e) represent the momentum distribution via the ridge-bypassing (red) and over-the-ridge (black) dissociation pathways. (f) The branching ratios between the ridge-bypassing [$\sigma_{anion}(^2A')$] and the over-the-ridge [$\sigma_{anion}(^2\Pi)$] pathways.

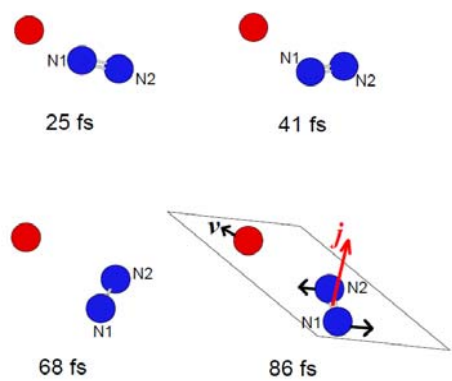

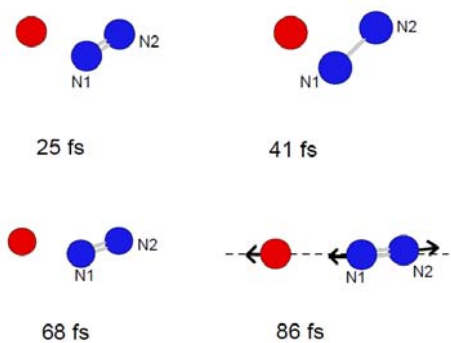

**FIG. 4** Eight frames of two sample trajectories with the low initial excess energy of 0.01 eV (a) and the high energy of 1.00 eV (b). Oxygen atom is in red and nitrogen atom is in blue.

*Supplementary Information of*

# Imaging the Indirect Dissociation Dynamics of Temporary Negative Ion: $N_2O^- \rightarrow N_2 + O^-$

*Lei Xia, Bin Wu, Hong-Kai Li, Xian-Jin Zeng, and Shan Xi Tian\**

Department of Chemical Physics, Hefei National Laboratory for Physical Sciences at the Microscale, University of Science and Technology of China, Hefei, Anhui 230026, China

\* E-mail: sxtian@ustc.edu.cn

## Experimental Method

Our newly developed anion velocity image mapping apparatus has been described in detail elsewhere[1] and described as the following figure S1. In brief, an effusive molecular beam is perpendicular to the pulsed low-energy electron beam which is emitted from a homemade electron gun; these low-energy electrons are collimated with the homogenous magnetic field (15 – 20 Gauss) produced by a pair of Helmholtz coils (diameter 800 mm). Thermal energy spread of the electrons is about +/-0.3 eV. The anionic fragment yields of the DEA are periodically (500 Hz) pushed out from the reaction area then fly through the time-of-flight (TOF) tube (installed along the molecular beam axis, the total length is 360 mm). Ten electrodes of the TOF mass spectrometer are in charge of the spatial ($2 \times 2 \times 2$ mm$^3$) and velocity ($\Delta v/v \leq 2.5\%$) focusing of the anions. The anionic fragments produced during one electron-beam pulse will expand in the three-dimension space and form a Newton sphere. Finally they can be detected with a pair of micro-channel plates (MCPs) and a phosphor screen. The three-dimension O$^-$ momentum distributions were directly recorded with a CCD camera using the time-sliced imaging technique,[1,2] namely, a detection time-gate is realized with a high voltage pulse (50 ~ 60 ns width) added on the rear MCP. The magnetic field used for collimating the electron beam cannot distort the ion image, but shift the image as whole. More information about the influence of the magnetic field can be found in ref. (1) and the reference cited therein.

Due to the much lower cross sections of DEA process with respect to those of ionization or excitation, it is still a challenge for application of the electron monochromator in the present apparatus (but we are striving now!). The energy spread of the electron beam was estimated by measuring the threshold ionization values of some atoms. The molecular temperature in an effusive molecular beam was about dozens of K. The target molecular rotations are ignored in the experimental data analysis.

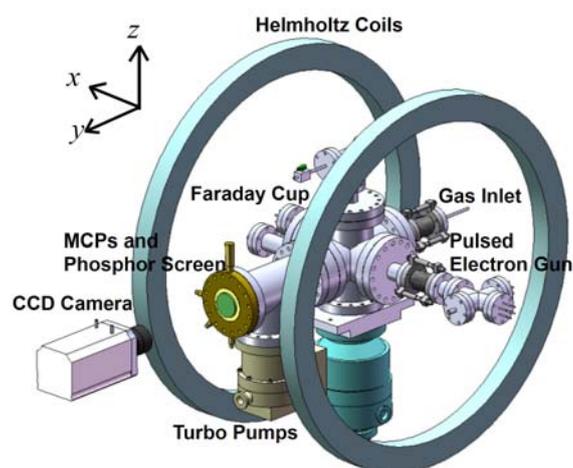

Figure S1. Scheme of our anion velocity image mapping apparatus

## Theoretical Methods

The complete-active-space self-consistent-field (CASSCF)[3] calculations were performed to plot the PES of the TNI system $N_2O^-$. In the calculations, eleven electrons and ten molecular orbitals (including five doubly occupied orbitals, one singly occupied orbital, and four virtual orbitals) were used to construct the active space. In the present case, only the lowest state ($^2\Pi$ with the linear structure and $^2A'$ with the bending structure) of $N_2O^-$ is considered, the basis set 6-311G(d) is good enough to describe the valence orbitals. Usually, more polarization and diffuse functions should be supplemented in the basis set for the excited states, in particular, of the negatively charged species.

The *ab initio* molecular dynamics (MD) simulations were performed with the atom-centered density matrix propagation [ADMP-B3LYP/6-311G(d)] method. In the ADMP approach,[4] the one-electron density matrix is expanded in an atom-centered Gaussian basis and is propagated as electronic variables. In comparison with the traditional semiempirical/MM Born-Oppenheimer dynamics. ADMP is much more efficient and can be combined with *ab initio* methods regarding Hatree-Fock and pure or hybrid density functional theory. ADMP can treat all electrons of the system in quantum chemistry scheme without resorting the pseudopotentials unless so desired and control the deviation from the Born-Oppenheimer surface precisely without the resulting mixing of fictitious and real kinetic energies. The normal Car-Parrinello (CP) method[5] treats the electrons separately, the core electrons with the pseudopotentials, and the valence electrons with the plane wave bases. Another important advantage of ADMP is that the charged system can be treated as easily as the neutral, owing to the correct physical boundary conditions combined with the atom-centered functions. CP MD usually needs to replicate cells in order to impose three-dimensional periodicity, leading to the troublesome for the charged system. In the present simulations, the initial structure of $N_2O^-$ is the same as the equilibrium structure of the neutral. Considering the vertical electron attachment, the structural propagation is initiated in the Franck-Condon region of the promotion from the neutral to the TNI. The adiabatic ADMP-MD simulations were performed with time step of 0.2 fs, the initial or system kinetic energy of 0.01 or 1.00 eV, and fictitious electronic mass of 0.1 amu bohr$^2$.

All calculations were performed with quantum chemistry package GAUSSIAN 03.[6]

Above electronic structure calculations and simulations could provide us a vivid dynamic picture of the indirect dissociation of TNI $N_2O^-$, but the electron autodetachment due to the coupling with the electron continuum background[7] is not considered. The autodetachment is another important process competing with the dissociation of TNI. The more sophisticated quantum theory method (see ref.7 and 8) is required. However, the present electronic structure calculations could be applicable since the complex PES ($E_r + i\Gamma/2$) could be simplified as its real part ($E_r$) in asymptotic region (fragments). The virtual $\Gamma$ (representing the lifetime of TNI) value would decrease rapidly when the TNI is decomposed into the stable fragments.

## Fitted Parameters:

Using eq.(2), the phase lags with respect to *s* partial wave for $^2A'$ resonant state are $\delta_1$ (*p* partial wave) = 2.383 rad and $\delta_2$ (*d* partial wave) =0.939 rad (at 0.70 eV), $\delta_1$ = 2.503 rad and $\delta_2$ = 1.296 rad (at 1.90 eV), $\delta_1$ = 2.601 rad and $\delta_2$ = 1.312 rad (at 2.25 eV), and $\delta_1$ = 2.746 rad and $\delta_2$ = 1.940 rad (at 2.60 eV); the phase lags with respect to *p* partial wave for $^2\Pi$ resonant state are $\delta_3$ (*d* partial wave) = 1.981 rad and $\delta_4$ (*f* partial wave) = 1.982 rad (at 0.70 eV), $\delta_3$ = 1.868 rad and $\delta_4$ = 1.916 rad (at 1.90 eV), $\delta_3$ = 2.096 rad and $\delta_4$ = 0.716 rad (at 2.25 eV), and $\delta_3$ = 2.794 rad and $\delta_4$ = 1.161 rad (at 2.60 eV). These phase lags arise from the different influences on each partial wave of the incident electron by the interaction potential of the target and represent the scattering partial-wave interference effect on the angular-resolved cross sections.[7] The weights of each partial-wave amplitude are in the ratio $a_0 : a_1 : a_2 : b_0 : b_1 : b_2$ of 1.00 : 0.61 : 0.21 : 0.23 : 0.80 : 0.00 (at 0.70 eV), 1.00 : 0.57 : 0.31 : 0.77 : 0.92 : 0.04 (at 1.90 eV), 1.00 : 0.57 : 0.29 : 0.75 : 0.72 : 0.45 (at 2.25 eV), and 1.00 : 0.76 : 0.26 : 0.80 : 0.85 : 0.48

(2.60 eV). One can find that the predominant scattering amplitudes are $s\Sigma(^2A')+d\Pi$ at the incident energies of 0.70 and 1.90 eV, while $s\Sigma(^2A')+p\Pi+d\Pi$ at 2.25 and 2.60 eV.